\documentclass[]{aa}  
\DeclareMathAlphabet\mathbfcal{OMS}{cmsy}{b}{n}
\usepackage{bm}
\usepackage{graphicx}
\usepackage{savesym}
\usepackage{amsmath}
\usepackage{amssymb}
\usepackage{txfonts}
\usepackage{time}
\usepackage[usenames]{color}

\usepackage{natbib}
\bibpunct{(}{)}{;}{a}{}{,}

\def\setid#1,v #2 #3/#4/#5 #6 #7 #8.{(\textit{CVS version #2, checked in by #7 on #3-#4-#5, #6 UT})}

\begin{document}

\newcommand{\comment}[1]{{\color{red} #1}}
\newcommand{\todo}[1]{{\color{red} #1}}
\newcommand{\fix}[1]{{#1}}
\newcommand{\lf}[1]{{#1}}
\newcommand{\lff}[1]{{#1}}
\newcommand\mymathbf[1]{{\bm{#1}}}
\def\abs#1{|#1|}
\newcommand\langcorr[1]{{\bf#1}}
\newcommand\replyoneadd[1]{{#1}}
\newcommand\replyonedel[1]{}
\bibliographystyle{aa}
\frenchspacing
\title{Image restoration of solar spectra}
 
\author{M.\ van Noort }

\date{Draft:  \now\ \today\ \setid $Id: ms.tex,v 1.2 2015/01/14 09:29:41 noort Exp $.}
                  
\authorrunning{M. van Noort}

\offprints{M. v. N.: \email{vannoort@mps.mpg.de}}

\institute{Max-Planck Institute for Solar System Research,
  Justus-von-Liebig-Weg 3,
  D-37077 G\"ottingen, Germany}

\abstract
{When recording spectra from the ground, atmospheric turbulence causes degradation of the spatial resolution.}
{We present a data reduction method that restores the spatial resolution of the spectra to their undegraded state.}
{By assuming that the point spread function (PSF) estimated from a strictly synchronized, broadband slit-jaw camera is the same as the PSF that spatially degraded the spectra, we can quantify what linear combination of undegraded spectra is present in each degraded data point.}
{The set of equations obtained in this way is found to be generally well-conditioned and sufficiently diagonal to be solved using an iterative linear solver. The resulting solution has regained a spatial resolution comparable to that of the restored slit-jaw images.}
{We have developed a new image restoration method for the restoration of ground-based spectral data over a large field of view. The method builds on the PSF information recovered by the MOMFBD code and typically reaches a spatial resolution comparable to that of the broadband slit-jaw images used to recover the PSF.}


\maketitle

\section{Introduction}
One of the main advantages of observational solar physics\replyonedel{ and many other forms of observational astronomy} has traditionally been the ability to resolve the individual features on the surface of the Sun. Many advances in solar physics have been, at least in part, due to advances in observational capabilities, either in sensitivity, or in resolution in the spatial, spectral, or time dimensions.

Increases in the throughput and optical performance of ground-based solar instrumentation (e.g., SST/CRISP, NST/VIS and DST/IBIS), all supported by adaptive optics systems, have produced a dramatic increase in the quality and time resolution of the available image data in the last two decades. This is further enhanced by an equally impressive increase in computing power in recent times that has made the use of post-facto image restoration techniques (e.g., SPECKLE, [MO]MFBD) routine practise. Consequently, a solar telescope is now considered competitive only when it is able to reach a spatial resolution comparable to its theoretical diffraction limit on a number of days during each observing season, covering a continuous, uninterrupted period of at least an hour, when image restoration techniques are used to restore the raw data.

For a number of reasons, however, advances in the acquisition of data of high spectral resolution have not been quite as impressive. Although high-spectral-resolution maps obtained in an optically stable environment are an excellent way to study the solar photosphere, as was demonstrated convincingly by the spectropolarimeter onboard the Japanese Hinode satellite \citep{2007SoPh..243....3K}, progress in obtaining such data from the ground has been significantly less convincing, where the most significant advance has been the use of advanced solar adaptive optics systems \citep{1989A&A...224..351V,1999ASPC..183..222R,2007PASP..119..170D,2016SPIE.9909E..0XS} in combination with traditional long-slit spectrographs. While the improvements made by adaptive optics systems to spectrograph data are noteworthy and considerable \citep{1996A&AS..115..367C}, the spatial resolution of such observations still does not compare favorably to that of imaging data restored using image restoration techniques \citep{2001A&A...367.1011H}, primarily due to the long exposure time typically needed to compensate for the high spectral resolution of the data, but also due to the removal of one of the spatial dimensions by the spectrograph slit, necessary to obtain the high spectral resolution.

While the long exposure time leads to a smearing of the solar scene, which is blurred and moved around on a short time scale due to atmospheric turbulence, the absence of the full spatial information from the data prevents the direct application of an image restoration technique to the data. Although \cite{1995A&AS..110..565K} and later \cite{2000SoPh..194...35S} already attempted to use a parallel slit-jaw camera to deconvolve the spectra using the speckle technique, their results did not capture much attention and remained virtually unused for over 20 years. \cite{2011A&A...535A.129B} continued their efforts by attempting to estimate the straylight contamination in very general terms, enhancing the contrast, but not dramatically improving the spatial resolution.

In this paper, we explore a new method that uses wavefront information obtained with the MFBD technique from a slit-jaw camera to restore strictly simultaneously recorded spectral data.  Although developed independently, and not based on the speckle technique, the method uses a formulation similar to the one used in \cite{1995A&AS..110..565K} .

\section{Restoration of spatially degraded data}
Over the past decade, the restoration of spatially degraded data has become an indispensable tool in exploring the full potential of solar ground-based observational data. The use is now so widespread that new instruments are specifically designed with their benefits in mind, and it is not unusual to work exclusively with restored data, without ever looking at the original, degraded data.

While this is true for imaging instruments, the same cannot be said for spectrographic instruments, where progress in developing a spectral image restoration technique has been hampered by the absence of spatial information in the direction perpendicular to the slit. In this paper, we explore the possibility of restoring spectrograph data, in the presence of detailed information on the shape of the PSF that degraded the data before it was sampled by the spectrograph slit.

To obtain such information, a so-called slit-jaw camera is used to observe the degraded image on the slit, in a wavelength band containing the recorded spectra, strictly synchronized with the spectral camera.

\subsection{Restoration of image data}
The restoration of spatially degraded image data can be largely divided into two main streams of techniques: the direct inversion methods and the forward modeling techniques. While the former are typically fast and robust, the latter are more careful, flexible, and informative, which explains their peaceful co-existence in observational solar physics in the form of the Speckle and [MO]MFBD post-processing options for many years.

Speckle image restoration directly recovers the Fourier phase of the restored images from the data, and is therefore a method that clearly belongs in the first category. The reconstruction of the phase is possible due to the specific form of the image degradation induced by atmospheric turbulence, which can be shown to preserve certain properties of the Fourier phase of the image \citep[see for instance][and many more]{1970A&A.....6...85L,1974ApJ...193L..45K,1983ApOpt..22.4028L,1992A&A...257L...4D,1993A&A...268..374V}. The Fourier amplitudes are recovered using assumptions regarding the statistics of the turbulence, and how the the performance of the adaptive optics system of the telescope affects these. The result of this class of methods is the most likely estimate of the undegraded image.

Multi-Frame Blind Deconvolution (MFBD, \cite{2002SPIE.4792..146L}), on the other hand, firmly belongs in the realm of forward modeling and optimization methods. This particular method is based on the assumption that the image degradation is solely due to a phase error in the incoming wave of light, and that, given an estimate of this wavefront error for a set of individually degraded images, it is possible to find the most probable undegraded image using only the data and the wavefront errors. 
Degradation of this undegraded image estimate then produces an artificial dataset that can be subtracted from the observed data and used to drive the wavefront error estimate to an optimum, given by that set of values that minimizes the differences between the observed and artificial data. The primary result of this method is an estimate of the \replyonedel{wavefront error}\replyoneadd{PSF for each image, described as the product of a pure wavefront error}; the restored image is merely a necessary by-product of the minimization procedure.

\subsubsection*{MFBD}
For the restoration of the slit-jaw data, we follow \cite{2005SoPh..228..191V} and define a transfer function $S_{\!j}$ for realization $j$, in terms of a wavefront error, which in turn is described in terms of a polynomial of wavefront modes $\phi_i$, with coefficients $\alpha_{i,j}$. Each $S_{\!j}$ can thus be parameterized by a set of several coefficients, which are optimized by minimizing the error metric
\begin{equation}
  \label{eq:L-mfbd}
  L(\alpha_{i,j}) = \sum_{u,v}\left[
  \sum_j^J \abs{D_{j}}^2 -\frac{\abs{\sum_j^J D^*_{j} \hat S_{\!j}}^2}{\sum_{j}^J\abs{\hat S_{\!j}}^2 + \gamma}
  \right],
\end{equation}
that quantifies in Fourier space the sum of the square of the difference between the data $D_{j}$ and the restored object, degraded with the current estimated individual transfer functions $\hat S_{\!j}= S_{\!j}(\hat\alpha)$. In the usual case that the image data has a good signal to noise ratio, the regularization constant $\gamma$ may be set to zero. 

Equation \ref{eq:L-mfbd} makes use of the assumption of additive, Gaussian noise, in which case the common estimate of the Fourier transform of object $i$, $\hat F$, for a given set of transfer functions $S_{\!j}$ is given by \citep[see e.g.,][]{1996ApJ...466.1087P}
\begin{equation}
 \hat F= \frac{\sum_j^J D_{j} \hat S^*_{\!j} }{\sum_{j}^J\abs{\hat S_{\!j}}^2},
\end{equation}
in which the object is obtained by deconvolution of the average of a purposely degraded dataset, with the average over the autocorrelation function of the individual transfer functions. Since the data have already been degraded by the $S_{\!j}$, the additional degradation of the data with the complex conjugate of the $S_{\!j}$ results in a dataset that has been degraded with a PSF that is centered on the origin, and which can therefore be added to other such data, without further degrading the result.

We note that it is implicitly assumed that the transfer function $S_{\!j}$ is independent of the spatial coordinates, so that the convolution operator can be conveniently expressed as a multiplication in Fourier transformed space. While this is in reality not quite true, the assumption leads to a considerable reduction in computational effort, and \replyoneadd{for a 1m-class solar telescope} it is normally a good assumption over an area smaller than about 5"x5". In practice, an image is therefore divided in tiles no larger than approximately 5"x5", restored on a tile-by-tile basis, and then re-assembled to a single restored image\replyoneadd{, although this may become more problematic with the large 4m-class solar telescopes currently planned or under construction (DKIST, EST), where this area is expected to reduce to only 2"x2"}

\replyoneadd{The initial assumption made by the MFBD process, that the PSF is the result of a pure phase error of the wavefront entering the telescope pupil, is clearly only true if the image is recorded significantly faster than the time scale over which the wavefront changes. This so-called seeing freezing time scale is variable, dependent on the telescope diameter, and has been determined in the past to typically lie at or below 10ms \citep{1981siwn.conf..491T} for a 1m-class telescope. If this time is exceeded significantly, the assumption of the PSF being the result of a pure phase error is no longer valid, and the fitted wavefront cannot be unambiguously interpreted as a wavefront anymore, as it describes a PSF that has been produced by integration over time of several wavefronts, which is inconsistent with the formation model. Due to the non-linearity of the mapping from wavefront to PSF, the correspondence of PSF to wavefront is generally non-linear and not unique. The latter, however, can be addressed by providing additional constraints, such as phase diversity, that are effortlessly included in the formalism used here, and have been used to reduce or even remove the degeneracy completely (see for instance \cite{2000SPIE.4123...56C} or \cite{2007SPIE.6711E..03P}).}

\subsection{Restoration of slit-data}
The direct restoration of spatially degraded spectral data has \replyonedel{thus far}\replyoneadd{long} been considered unfeasible, since the image information required to constrain the undegraded image is missing in the direction perpendicular to the slit. To make matters worse, high-resolution spectra have a much lower photon flux than wideband filter images, and are therefore typically acquired using an integration time that exceeds the seeing correlation time by a considerable margin, so that even the assumption that the PSF can be calculated from the phase error of a single wavefront across the pupil cannot be used. \replyoneadd{The first serious attempt to deconvolve slit-spectrograph data was made by \cite{1995A&AS..110..565K}, followed up by \cite{2000SoPh..194...35S}, who used a parallel slit-jaw camera to deconvolve the spectra, using the speckle image restoration technique to obtain the atmospheric PSF. Due to technical limitations, these efforts did not capture much attention at the time, and were not followed up for nearly two decades.}

\replyoneadd{The problem has primarily been related to} the limited performance of science grade image sensors in the past, which was characterized by a relatively long readout time during which no photons could be collected, and a significant noise contribution to the sensor signal due to the process of reading it out. Major advances in recent years in the field of image sensors have allowed for a major increase in the readout speed, while simultaneously exposing, and with a reduction in the readout noise of almost an order of magnitude. It is thus possible to acquire an image in a short enough time to ``freeze'' the atmospheric seeing, without causing a significant degradation of the noise properties of the image due to readout noise.

In the following, we therefore assume that we are able to record spectra at or below the seeing correlation time, and without adding significantly to the photon noise. We further assume that we have a strictly synchronized camera, observing the part of the image that is not accepted by the spectrograph slit, and which we can use to obtain the PSF at every point around the slit using a broadband filter covering the wavelength range sampled by the spectrograph. 

If we now consider the spectrum from one point $(u,v)$ on the Sun, our knowledge of the fully space-variable PSF $\varphi(x,y)$ at every point in the field of view (FOV) allows us to write for the observed data
\begin{equation}
\delta_{i,x,y,u,v}=\varphi_{i,u,v}(x-u,y-v) s(u,v),
\end{equation}
so that the contribution of all points on the Sun to one data point is then given by
\begin{equation}
d_{i,x,y}=\sum_u \sum_v  \delta_{i,x,y,u,v} = \sum_u \sum_v \varphi_{i,u,v}(x-u,y-v) s(u,v),
\end{equation}
where we have assumed that the PSF $\varphi$ is the same for all wavelengths under consideration.

Taking the slit to be aligned exactly with the $y$-axis,\replyoneadd{ located at $x=x_{slit}$,} each data point thus constrains a sum over a region on the solar surface, which can be conveniently written as
\begin{equation}
d_{i,y,x_{slit}}={\bf a}_{i}^T \cdot {\bf s},
\end{equation}
where ${\bf a}$ contains the appropriate PSF coefficients and ${\bf s}={\bf s}(u,v,\lambda)$ is the undegraded three-dimensional solar scene. Every pixel in the spatial direction of a spectrum from a long-slit spectrograph yields one such equation for every exposure of the camera, resulting in\begin{equation}
{\bf J} \cdot {\bf s} = {\bf d},
\end{equation}
where each row of the Jacobian matrix ${\bf J}$ describes the contributions to each data point in the spectrum. 

Although we have assumed that the spatial PSF is independent of wavelength, this does not by itself decouple the wavelengths from one another, since the slit has a finite width, and the dispersion "smears" the slit across the spectrum, so that each data point actually contains contributions from a range of wavelengths. While this is trivial to write down, and equally trivial to solve for, it has significant implications for the numerical size of the problem, and we must therefore assume for now that the width of the slit is negligible compared to the spatial structure in the data, and negligible compared to the spectral structure. In that case, the coupling can be completely ignored, and we are left with an overdetermined but otherwise linear problem of moderate size.

As usual, the system must be solved by using the pseudo-inverse, which can be obtained by multiplying with ${\bf J}^T$ on each side, yielding the square system
\begin{equation}
{\bf A} \cdot {\bf s}\equiv {\bf J}^T \cdot {\bf J} \cdot {\bf s} = {\bf J}^T \cdot {\bf d} \equiv {\bf b}.
\label{eq:linsys}
\end{equation}

In principle this system has a number of variables equal to the size of the region that we want to restore - typically $10^7$ pixels - and is not sufficiently sparse to be handled in its entirety using readily available numerical resources. Under moderate to good seeing conditions, however, the radius of influence of the PSF is limited to a couple of arcseconds or so, so that we can deal with the problem in a segmented way, reducing the number of variables to approximately $10^4$ per segment.

\subsection{Conditioning the dataset}
\label{sec:condition}
To ensure that (\ref{eq:linsys}) is conditioned sufficiently well to be solvable, we must ensure that enough linearly independent datapoints are in the dataset. While we can rely on the seeing to provide this over time (i.e., sit and stare), it is unlikely that such data will put strong constraints on points at some distance from the slit. Moreover, the exact value of the PSF at large distances from the center is most likely not as reliable as near the center, leading to a larger statistical error in the equation, which can only be compensated for by adding more independent datapoints.

This suggests that it might be helpful to provide some assistance to the seeing in the generation of linearly independent realizations. One way of accomplishing this is by continually moving the location of the slit to different positions on the Sun, so that independent information over a spatially extended region is obtained.

This approach resembles that of a traditional spectral scan, with the notable differences that the assumption that a specific location on the Sun is observed is neither valid nor of importance, and that with the performance of current state-of-the-art Adaptive Optics systems, it is not needed or even useful to scan in discrete steps, since the actual observed position can only be deduced post-facto, and not imposed a priori, due to the stochastic nature of the residual seeing.

\section{Data reduction}
The method described in the previous section was implemented as a multithreaded C++ code, which provides sufficient performance for routine application of the method on a small number of modern workstations. The starting point for the reduction is the MFBD reduced slit-jaw images, obtained using the MOMFBD code. The data was reduced in bursts, with sufficient overlap in time to reliably calculate the alignment of the data across a large span of time, and with a dense grid of patches along the slit, to ensure \replyoneadd{that all anisoplanatic variations} are captured. 

\subsection{Data alignment}
Since the alignment of the data, that is, the location where the raw data is located relative to the solar coordinate system, is unknown, the MFBD process constrains the sum of all tilts in the dataset to vanish. This presents a problem, since \replyoneadd{the location of each restored image patch relative to another cannot be determined from the average tilt coefficients anymore.}\replyonedel{although the average tilt over an infinite period of time indeed must vanish on statistical grounds, bursts of data covering only several seconds have a residual error that cannot be neglected. }\begin{figure}[hbt]
  \begin{picture}(20,165)
    \put(0.00,0.00){\includegraphics[width=\columnwidth,angle=0]{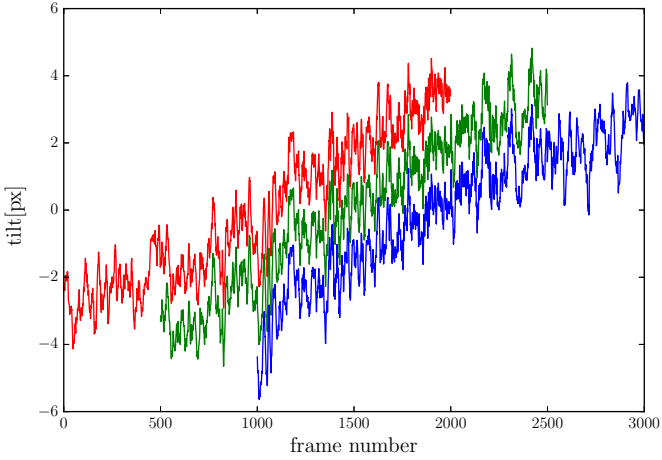}}
  \end{picture}
  \caption{Tilt component in the direction perpendicular to the slit, for three consecutive bursts of 10s worth of slit-jaw images, recorded at 200 fps. Although not identical, the coefficients of the overlapping frames are clearly similar, except for a global shift.}
  \label{fig:tt-align}
\end{figure}

\replyoneadd{Restoring a dataset spanning}\replyonedel{Averaging over} a longer period of time is not really possible, since solar evolution will cause the solar scene to change, thus \replyoneadd{compromising}\replyonedel{complicating} the determination of the tilt coefficient. We \replyonedel{must} therefore map the tilt across a longer period of time by comparing sections of data that are partially overlapping in time.

The mapping of one time section of the dataset to the next is accomplished by comparing the tip-tilt components of the overlapping part of two reduced bursts of data. The differences in the fitted tip-tilt coefficients for the overlapping exposures is, as expected, usually nearly identical for all overlapping data frames, with a residual that is much less than a pixel. If sufficient overlap was available, the average of this residual was found to be about $\frac{1}{100}$ of a pixel. We can thus accurately calculate the relative position of the frame of reference of the \replyoneadd{PSFs}\replyonedel{wavefronts} for each burst by averaging over the overlapping part of the burst. 

Figure~\ref{fig:tt-align} shows the tilt components in the y direction for three bursts of 2000 frames, offset in time by 500 frames. The difference between the two is nearly constant, and corresponds to the displacement of the solar scene in the time the two bursts are offset to each other; in this case the time it takes to record 500 frames.

\replyoneadd{We can use the alignment information obtained in this way to position each PSF in a {\em global} frame of reference, spanning the entire scan, instead of the arbitrary frame of reference of the dataset that the frames happened to be contained in. This global position in the scan is then used to calculate what part of the PSF falls onto the slit for each solar "pixel", and for each PSF in the scan. While in principle one could replace this alignment procedure with one based on cross-correlations of the restored images, using the tilts from the restorations completely avoids using the restored images, and appears to be less sensitive to restoration artifacts.}

\subsection{Calculating the PSF}
The MFBD restoration of the slit-jaw data produces as the primary result the wavefront error for each patch that was restored, for each data frame in the dataset. Although it is straightforward to calculate the PSF for each patch from this, when the PSFs of two neighboring patches are compared, they frequently show significant differences, even over sub-arcsecond distances. This presents an inconvenient inconsistency, as the assumption of a single wavefront error for each patch, made by the MFBD algorithm, is clearly not strictly valid. 
\begin{figure}[hbt]
  \begin{picture}(200,155) \put(0.00,0.00){\includegraphics[width=\columnwidth,angle=0]{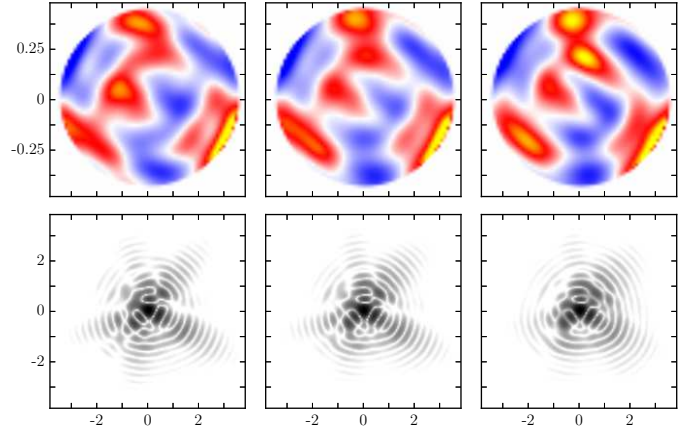}}
  \end{picture}
  \caption{Examples of a number of independent fits of the same wavefront. The resulting wavefronts are all very different, but all give rise to very similar PSF. The spatial scale of the PSFs is in arcseconds, while that of the wavefronts is in meters.}
  \label{fig:wavefronts}
\end{figure}

Addressing this problem thoroughly is beyond the scope of this paper, and we resort instead to determining the full spatial dependence of the PSF by considering them to be approximately correct, and continuously varying in space in a smooth manner, so that the full spatial dependence may be obtained by interpolation. Although by itself this is probably a fairly accurate approximation, the interpolation of the PSF is not a trivial task. 

Many of the changes between neighboring patches are indeed small, but involve a shift or rotation of the PSF, without altering its shape significantly. Direct interpolation in such a case will cause significant blurring and broadening of the PSF, which will probably cause significant errors in what is essentially a deconvolution.

The obvious solution, the spatial interpolation of the wavefront errors, exposes a major shortcoming of the MFBD process, in that not only can neighboring patches have totally different wavefront error estimates, even the same exposure, reduced as part of different bursts of data will generally yield a completely different fit to the wavefront. This is not an error in the algorithm, but rather a consequence of the strong level of degeneracy of the mapping from wavefront to PSF, as the PSFs yielded by differing wavefronts for the same patch of the same data frame are actually virtually indistinguishable.

Figure~\ref{fig:wavefronts} shows a set of fitted wavefronts for the same patch in the same dataframe, but reduced in a number of different bursts. Clearly, although the wavefronts are all different, the PSF they result in are all more or less the same. An obvious way to alleviate this problem would be to make use of a phase diversity channel, thus reducing the degeneracy.

The effect of interpolating the wavefront regardless of the large differences is generally that the interpolated wavefront is too flat, due to the inherent averaging of two different wavefronts, and thus the PSF is more compact than it should be. Since this creates more problems than the former, we resort to direct \replyoneadd{linear} interpolation of the PSF \replyoneadd{along the slit}, for which it was experimentally determined that a patch separation of only ten pixels was required to keep the blurring of the PSF \replyoneadd{below}\replyonedel{to} an acceptable level.

To take the width of the slit into account, we must add up contributions to several slit-jaw pixels to obtain the contribution to the light entering the slit. If the slit width is understimated, this results in artifacts in the direction of the slit due to reconstruction errors of seeing-induced fluctuations in the amount of signal that was accumulated for the different lines in the degraded image. Although the appropriate width is not critical, when selected correctly it is possible to almost completely eliminate all vertical stripes.

\subsection{Flatfield correction}
Since the observed area of the Sun is much larger than the same area in the focal plane of the telescope, the beam is slowly diverging, and thus the pupils for different areas on the solar surface are spatially separating with height. If the PSF for solar data is generated in a turbulent layer at some distance from the telescope entrance, where the pupils for different parts of the FOV are passing through different areas of turbulence, the primary effect of this is that the wavefront error induced by the turbulence is spatially varying. This spatial variation obviously also includes the tip-tilt components of the wavefront, resulting in a pattern of warping, compression, and stretching of the image, that is frequently observed, especially when the Sun is at low elevation.

These rapid continuous spatial variations of the PSF introduce some new and unexpected problems that require special consideration, starting with the description of the PSF as a continuously varying quantity. 
\begin{figure}[hbt]
  \begin{picture}(200,155) \put(0.00,0.00){\includegraphics[width=\columnwidth,angle=0]{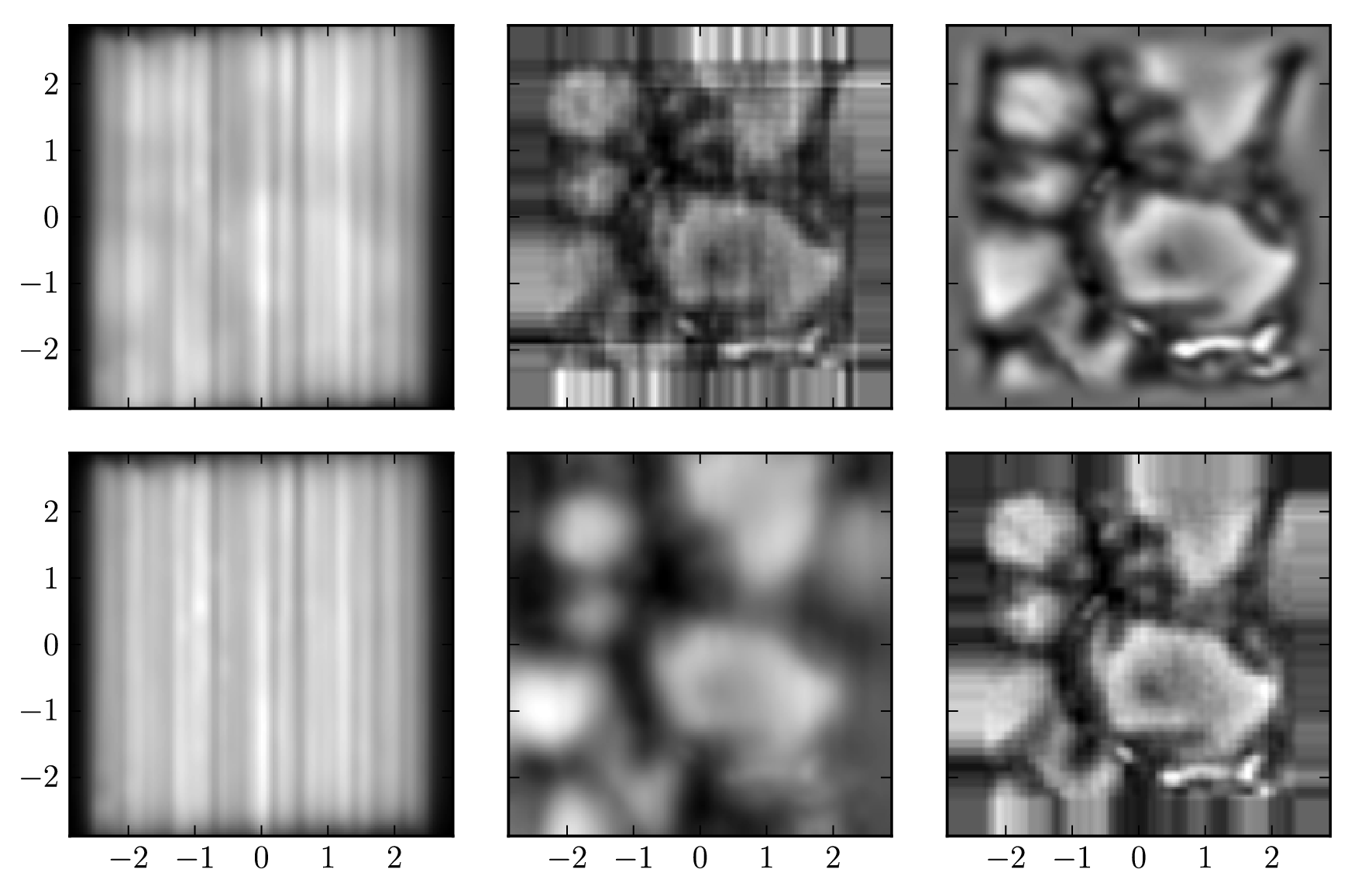}}
  \end{picture}
  \caption{\replyoneadd{Example, at an arbitrary continuum wavelength, of the need to correct for the unknown amplitude of the PSF. Top left: RHS data without a correction applied. Top center: restored monochromatic image without correction applied. Top right: restored slitjaw image. Bottom left: the flatfield. Bottom center: RHS with the flatfield correction applied. Bottom right: restored image with all flatfield corrections applied.} }
  \label{fig:flatfield}
\end{figure}
The warping and bending of the image itself can be accurately mapped and corrected for using this description, but it has the undesirable side-effect of redistributing the energy of the undegraded image inhomogeneously. The result is that while at first it would appear to be more accurate to allow the PSF to be a continuously varying quantity, it inevitably implies a continuously varying image scale, that is accompanied by a continuous variation of the image intensity. \replyoneadd{This effect is also observed in  MCAO research, where similar intensity fluctuations have been found \citep{2004SPIE.5490..617V}.}

In addition to this warping, the local seeing is moving the image randomly across the slit during the recording of the dataset, leading to differences in the amount of time the different image lines of the degraded image are sampled by the slit. These differences enter both the left hand side (LHS) and the right hand side (RHS) of (\ref{eq:linsys}), producing a RHS with strong fluctuations in the direction of the slit, as can be seen in the top right panel of Fig.~\ref{fig:flatfield}. 

The combination of these two effects produces artificial structure in the RHS that is so strong that it obscures the real image information almost completely. Since the dominant structure, the vertical stripes in Fig.~\ref{fig:flatfield}, are caused by the latter effect, which is also contained in the mapping operator ${\bf A}^T {\bf A}$, the solution of (\ref{eq:linsys}), shown in the top-middle panel of Fig.~\ref{fig:flatfield}, does not contain them to the same extent. However, a strong residual pattern of intensity fluctuations is clearly still visible in the restored image, making it appear noisy and obviously inferior to the image restored using the MFBD image restoration code. 

\begin{figure}[hbt]
  \begin{picture}(200,300)
    \put(35.00,0.00){\includegraphics[width=0.7\columnwidth,angle=0]{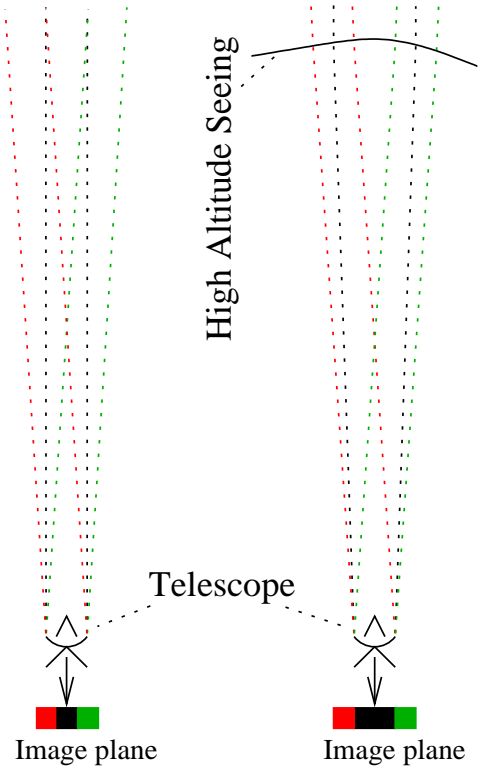}}
  \end{picture}
  \caption{Compensation effect of intensity change induced by differential seeing: the area sampled by the black rays is expanded in the image plane, distributing the light over a larger area and thus reducing the flux, but this is compensated for by a corresponding increase in the area of the wavefront that is collected by the telescope at the height of the seeing layer.}
  \label{fig:differential}
\end{figure}

The inability of the restoration process to eliminate the residual fluctuations is due to a \replyonedel{basic defect}\replyoneadd{shortcoming} in the description of the warping of the image, caused by the high-level seeing. Even though the warping is nearly always observed in real data, the intensity fluctuations that it is supposed to produce are not, although why that would be the case is unclear. The reason for their absence is illustrated in Fig.~\ref{fig:differential}. What looks like a continuous variation of the local tip and tilt components, on a larger scale amounts to a focus term, that is, an additional lens at high altitude that is positioned in front of the telescope. The effective area of the incoming wavefront sampled by the telescope is thus increased by the same fraction as the image scale on the detector, so that the total amount of light per pixel remains unchanged. 
 Because the current formalism describes only the change in the image scale, but not the change in the telescope \replyoneadd{aperture}\replyonedel{area}, fluctuations are resulting in the RHS and in the LHS of (\ref{eq:linsys}). While it is not difficult to find the correct position of an image pixel using image restoration, it is another thing entirely to produce from that the image scale, as this requires access to the undistorted image. Alternatively, one could assume that the average of the distortions vanishes over time, but the time scale on which this becomes a valid assumption can be very long compared to the solar evolution time scale, and it is an invalid assumption if the telescope is systematically moved across the solar surface, as is the case for a spectral scan, meaning that another way to deal with the intensity fluctuations needs to be found.

A solution can be found by making the assumption that if the Sun would present us with a featureless image, the dataset would also contain no structure. We can thus apply the operator ${\bf A}^T$ to the assumed ``proper'' featureless dataset, yielding the RHS that would have resulted in that case. The result is shown in the bottom left panel of Fig.~\ref{fig:flatfield}, and contains both real intensity fluctuations, caused by seeing induced global image motion, as well as artificial, ``warping'-induced ones.  Since there is insufficient information to separate these two contributions, we have no choice but to correct the RHS for both effects, by dividing the RHS by the artificial ``flat'' RHS. The result, shown in the bottom middle panel of Fig.~\ref{fig:flatfield}, no longer shows any obvious image fluctuations, and a blurred, averaged image of the scanned area can be discerned instead. 

Since we applied corrections for all seeing induced intensity fluctuations to the RHS of (\ref{eq:linsys}), we must now apply those corrections also to the LHS. Unfortunately, a correction of ${\bf A}$ for the individual PSFs cannot be calculated from first principles, since the fluctuations are, at least in part, the result of the unknown spatial amplitude variations of each individual contributing PSF. We therefore proceed by applying ${\bf A}^T{\bf A}$ to an image with a value of unity everywhere, to obtain the RHS that would have been obtained if that were the true image. We again make use of the assumption that if a featureless image were observed, a featureless dataset would be obtained, and calculate the corrected operator by dividing each row by the corresponding artificial RHS value. 

The solution of the corrected system is shown in the bottom right panel of Fig.~\ref{fig:flatfield}. Clearly, the result closely resembles the restored image in the top right panel, shows no signs of stripes or intensity artifacts, and even appears to rival the restored image for contrast and resolution.
\begin{figure*}[hbt]
  \begin{picture}(200,520) \put(0.00,0.00){\includegraphics[width=\textwidth,angle=0]{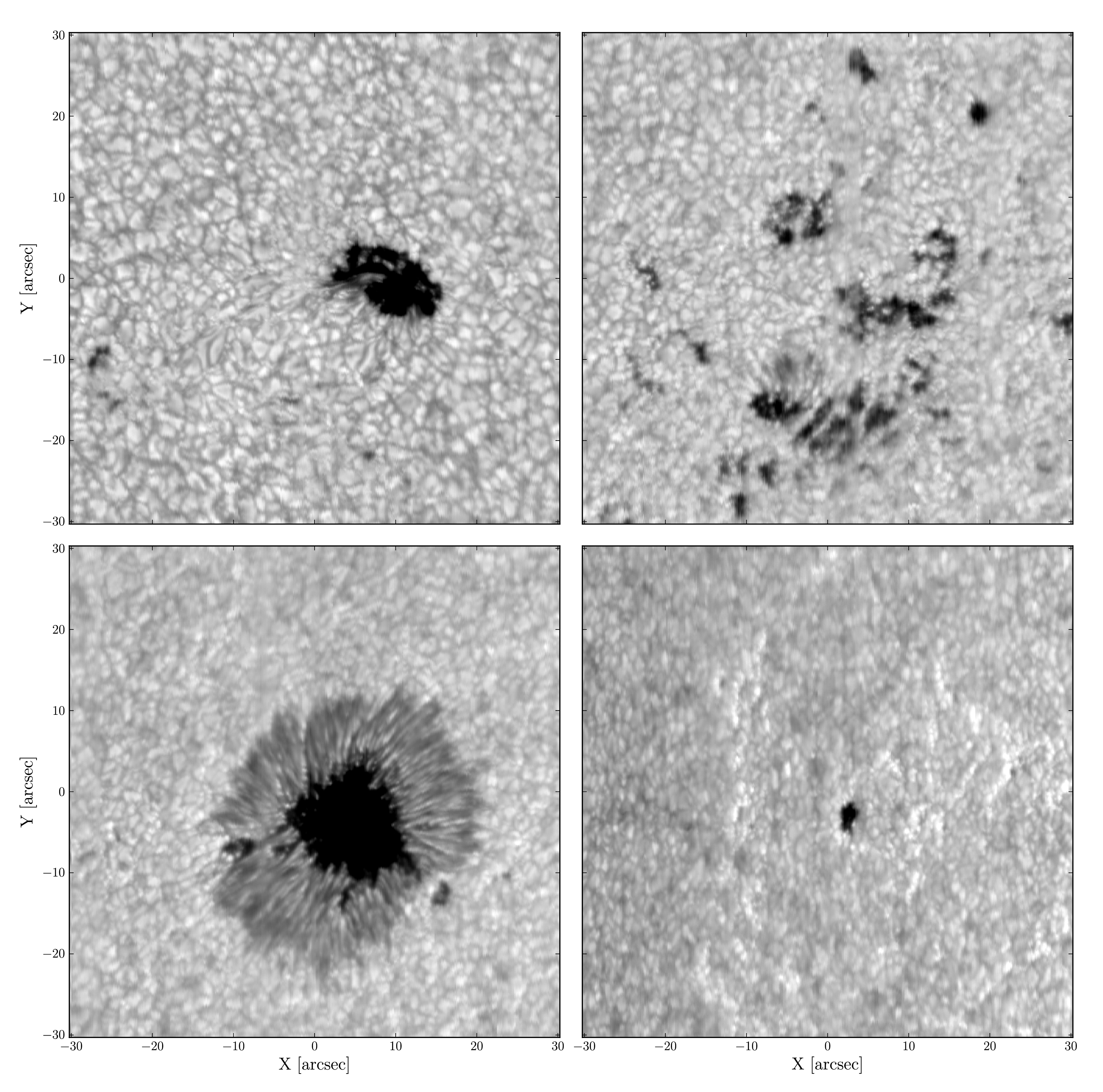}}
  \end{picture}
  \caption{\replyoneadd{Continuum images extracted from four unprocessed scans, obtained by directly summing 40 raw spectra for each vertical image line, each spanning a period of about 400ms, thus covering a period of approximately 600s for each scan. Top left: newly emerging AR at disk center, Top right: trailing plage of AR12436, located at $\mu$=0.82 Bottom left: Main spot of  AR12436 at $\mu$=0.69, Bottom right:  a small plage region towards to the east limb, at $\mu$=0.53. }}
  \label{fig:overview}
\end{figure*}

\subsection{Solving the linear system}
Due to the size of the system to be solved, an iterative method was selected. One of the main characteristics of this type of solver is that to compute the correction, only an approximation to the full operator is used. Although this makes the inversion of the approximate operator fast, it can lead to large inaccuracies in the correction, since taking only the diagonal implicitly assumes that the current defect is solely due to a local error in the solution. If the problem is not very local in reality, an overcorrection is applied that grows exponentially with every iteration, and the solution diverges. 

In the method employed here, only the diagonal was selected for the approximate matrix to be inverted, leading to an intrinsically unstable scheme, that must be damped with an appropriate factor to converge. However, this method has the advantage that if some areas of the undegraded image are not constrained sufficiently by the data, they can be set to a reasonable value, without jeopardizing the convergence of the constrained regions, as would be the case if a solution by direct matrix inversion was attempted.
\begin{figure*}[hbt]
  \begin{picture}(200,520) \put(0.00,0.00){\includegraphics[width=\textwidth,angle=0]{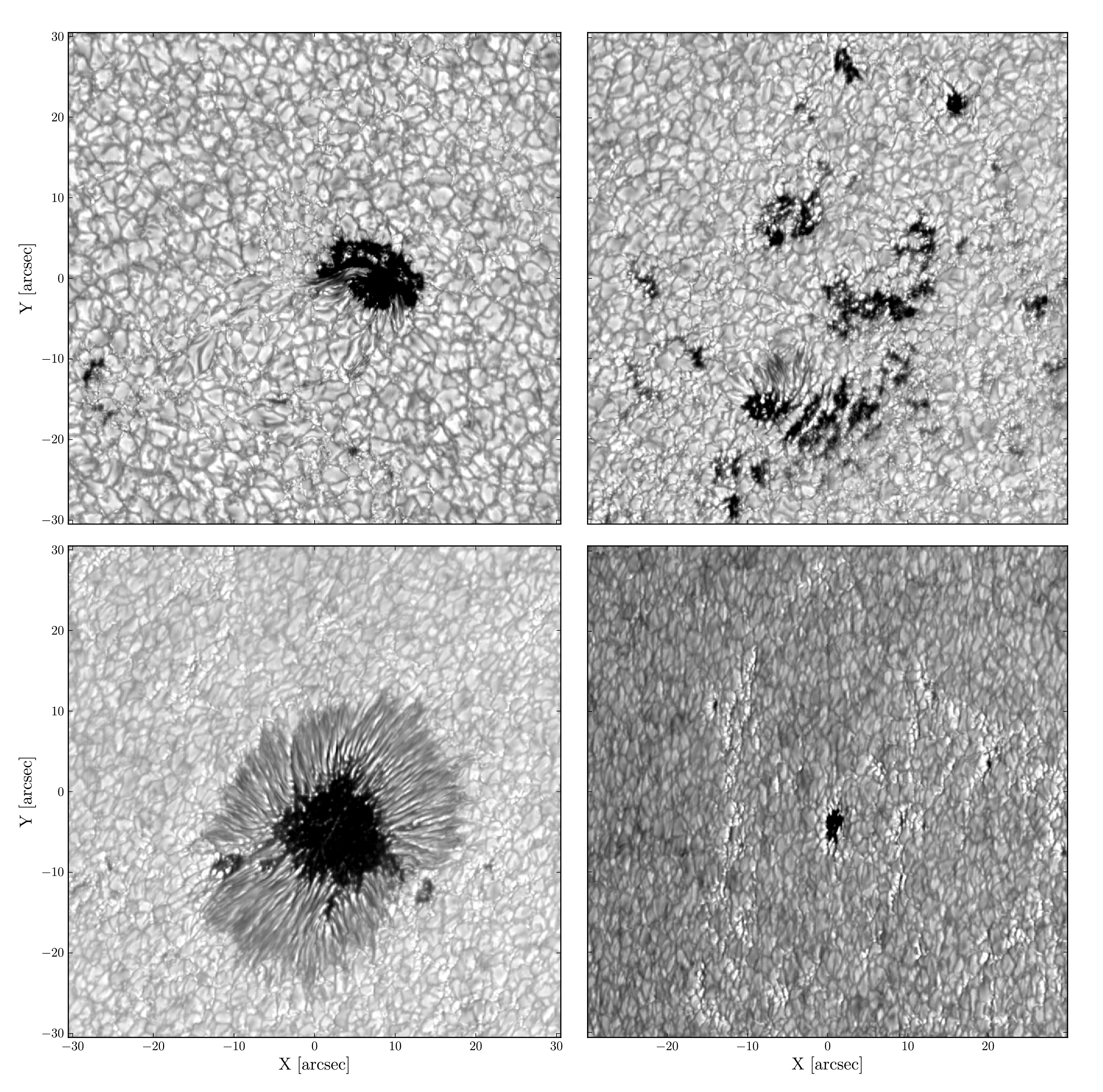}}
  \end{picture}
  \caption{\replyoneadd{Restored continuum images of the four scans shown in Fig.~\ref{fig:overview}, restored using the spectral restoration method described in this paper. Top left: newly emerging AR at disk center, Top right: trailing plage of AR12436, located at $\mu$=0.82 Bottom left: Main spot of  AR12436 at $\mu$=0.69, Bottom right:  a small plage region towards to the east limb, at $\mu$=0.53.}}
  \label{fig:overview_rec}
\end{figure*}

The convergence rate can be optimized by ``damping'' the correction with an estimate of the true local contribution. The smaller this is, the stronger the damping needs to be to make the solution converge. In the current problem, this implies that if the seeing is not so good, a stronger damping factor is required than when the seeing is excellent, taking more iterations to converge.

One of the most important disadvantages of using only the diagonal of an operator to calculate an approximate correction is that the correction is injected into the solution at the highest frequencies. This has the undesirable consequence that the larger the spatial scale of an error, the more iterations it takes to propagate across the FOV. We therefore employ a variation on a multigrid strategy, where the defect of an approximate solution is calculated, coarsified to a lower resolution and then approximately solved, resulting in a considerable improvement in the convergence rate of the solution.

\replyoneadd{The inversion process itself is unregularized, so that it needs to be terminated after a to-be-determined number of steps. This number was set experimentally, by evaluating the power at the high frequencies in the restorations. The number of iterations was set to the point where the power spectrum showed an increase in power for the highest frequencies, typical for  deconvolved data. This rather arbitrary and unsatisfactory situation can perhaps be formalized using a proper regularization strategy, but a suitable formulation for this process has not yet been finalized.}

\section{Observations}
To demonstrate the method, we observed a number of targets at the Swedish Solar Telescope (SST) on 26 October 2015. The seeing was moderate, with an estimated Fried parameter $R_0$ of about 10cm, over a period of about 2 hours in the morning, between 9:00 and 11:00 UTC.
\begin{figure*}[hbt]
  \begin{picture}(200,25) \put(0.00,0.00){\includegraphics[width=\textwidth,angle=0]{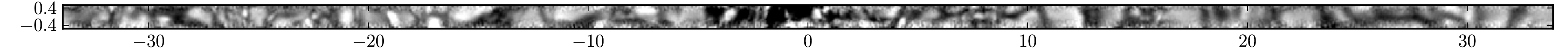}}
  \end{picture}
  \caption{Restored spectra of 23 seconds of data obtained without moving the 0.11" wide slit. Horizontal and vertical axes are in arcseconds. Although the AO and CT were both in closed loop operation, the variation of the PSF in the 23 seconds covered by the data was sufficient to recover the spectra over a 1" wide strip. The highest S/N of about 500 was found to be concentrated in the central 4 pixel rows, whereas the S/N in the outer few pixel rows is approaching unity, which is clearly visible in the image.}
  \label{fig:sit-n-stare}
\end{figure*}

The data were recorded with two large-format Jai SP20000 CMOS cameras, one of which was observing the region of 6297-6304\AA, containing three FeI lines, a TiI line and several other weak lines, as well as four telluric $O_2$ lines. The second camera was installed as a slit-jaw camera, viewing the slit plate through a 1:1 telecentric re-imager at a wavelength of $6302 \pm\ 4.4$\AA. In addition, two neutral density (ND)
filters were needed to reduce the light level by a factor of 100, to allow the exposure time of the slit-jaw camera to equal that of the spectral camera. Both cameras were windowed down to a region of 2500x1100 pixels, to allow them to achieve a frame rate of 100fps, at which rate they exposed continuously with an exposure time of about 10ms, a generic value for the time scale above which motion-induced blurring generally starts to outcompete the diffraction limit of a large-aperture telescope \citep{1981siwn.conf..491T}.

The spectral camera was aligned with the x-axis parallel to the slit, leaving the vertical direction for the spectral dimension. The slit-jaw camera was aligned as carefully as possible with the x-axis parallel to the slit, to simplify the data reduction, and to maximize the length along the slit for which the PSF could be calculated. Any image information away from the slit is not useful for the reconstruction, but it was kept as equally large as the spectral dimension of the spectral camera, ensuring strictly synchronous operation. Both cameras were externally triggered using a signal generator.
The movement of the slit across the Sun was controlled by the tip-tilt correction software of the AO system in the x-direction of the tip-tilt system. This did not exactly line up with the slit direction, but made an angle of a few degrees with it instead; this did not really present a problem, since the position is determined from the slit-jaw data, and is not required to be precise, as described in Sect. \ref{sec:condition}. The slit was moved by 0.01" at a rate of ten steps per second, which should have amounted to 0.1"/s. From the restoration, a rate of 0.144"/s was measured instead, suggesting that this process was not accurate by a considerable margin for as of yet undetermined reasons.
To get an impression of the data quality, for four scans, the spectral data was summed over 40 consecutive frames, covering a period of 400ms, not an unusual exposure time for spectrographic data. In this period of time, a distance of 0.058" is covered by the scan, which is close to critical sampling, that is, half of the diffraction limit of 0.13" of a 1m telescope at a wavelength of 6300\AA, and about twice the pixel size of the camera. 

Although it may seem like an inappropriate way to represent the data, since the slit was actively moving across the image during the exposure, from the MOMFBD process the motion of the image over this period of time induced purely by seeing was determined to have an RMS value of some 2 pixels in both the x and the y directions. On this scale, the blurring induced by the systematic drift of the scan is completely negligible, and the summed spectra are in fact a faithful representation of the data quality that would have been obtained if synchronized, discrete steps of 0.058"  had been taken every 400ms, or 80 frames in Fig.~\ref{fig:tt-align}.

Figure~\ref{fig:overview} shows a number of scans, binned vertically by a factor two, to obtain a nearly square pixel of 0.060"x0.058". The scans cover an area of approximately 60"x60", and show clearly that the seeing was fairly constant, but not excellent, with occasional moments of somewhat degraded seeing conditions, especially in the center of the top-right scan.

The scans cover a selection of active regions, at a variety of angles on the solar disc. The top left scan was a newly emerging active region, that emerged near disc center, the top right and bottom left scans are the trailing and leading polarities of AR12436, located at $\mu$=0.82 and $\mu$=0.69 respectively, and finally the bottom right scan is of a small plage region towards to the east limb, at $\mu$=0.53. 

\section{Results}
The data was first corrected for dark current, and multiplied with a gain correction table, derived from a large number of frames obtained with the telescope moving in a circle around the center of the solar disc. The slit on the slit-jaw camera was masked out and linearly interpolated in the direction perpendicular to the slit. The slit-jaw frames were then organized in bursts of 500 frames, spanning a period of 5 seconds, and reduced using the MOMFBD code \citep{2005SoPh..228..191V}, using a patch size of 96 pixels (2.9") and 44 orthogonalized Zernike modes. The PSFs obtained in this way were then processed using the method described above, yielding the results shown in
\begin{figure}[ht]
  \begin{picture}(200,237) \put(0.00,0.00){\includegraphics[width=\columnwidth,angle=0]{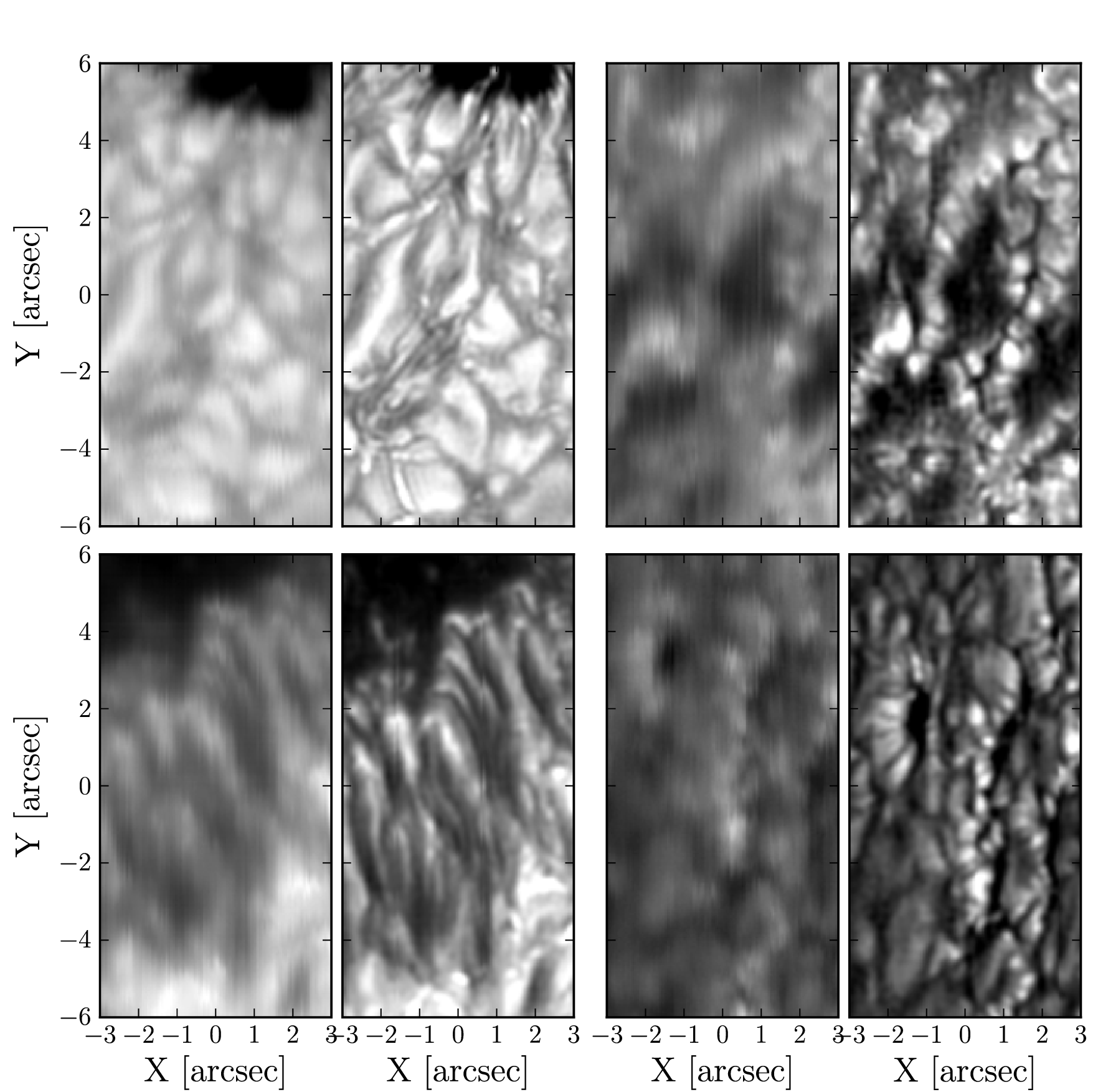}}
  \end{picture}
  \caption{\replyoneadd{Comparison of detailed sections of each scan shown in Figs. \ref{fig:overview} and \ref{fig:overview_rec}, in the order they appear in those Figures. The wavelengths are all in the continuum, with the exception of the top-right and the bottom-left panels, where the red wing of the  FeI 6302.5 line was selected to present examples that have a more favorable contrast. }}
  \label{fig:details}
\end{figure}
Fig.~\ref{fig:overview_rec}. 
The first, striking observation is that the seeing-variation-induced blurring, obviously visible in the summed scans, is not visible in the restored scans. In addition, the contrast is significantly enhanced.

\subsection{Monochromatic image data}
On the spatial scale of \langcorr{Figs. \ref{fig:overview} \& \ref{fig:overview_rec}}, the difference between restored and unrestored data is already clearly visible, with enhanced contrast and an increased brightness of magnetic elements. In particular, the blurred vertical stripes, clearly visible in several places in the unrestored scans, caused by persistent degradation of the seeing conditions over periods of tens of seconds at a time, are apparently absent from the restored scans. 

The difference becomes more obvious when we zoom in to an image scale where we can discern the smallest image details present in the data. A specific detail was selected from each scan, where these differences are most obviously visible, and shown, paired with the corresponding area from the unrestored scan, in Fig.~\ref{fig:details}. Although some of the raw scans could by themselves be described as being of reasonable quality, they do not compare favorably to their restored counterparts in terms of spatial resolution and consistency. 

From \langcorr{Figs. \ref{fig:overview} \& \ref{fig:overview_rec}, and from the top right scan in particular}, it is also clearly visible that the raw scan, that relies on the scan rate as an absolutely reliable horizontal coordinate, has an image scale that clearly differs from that of the restored scan. Not so clearly visible in the scans is that the scan direction of the tip-tilt mirror was not exactly perpendicular to the slit, causing the raw scans to be skewed. Since the restoration process tracks the true position of the slit on the Sun, not only are errors due to the scanning direction removed, but also deformations caused by rotation of the solar image on the table can be corrected.

\subsection{Spectral data}
\begin{figure}[ht]
  \begin{picture}(200,150) \put(0.00,0.00){\includegraphics[width=\columnwidth,angle=0]{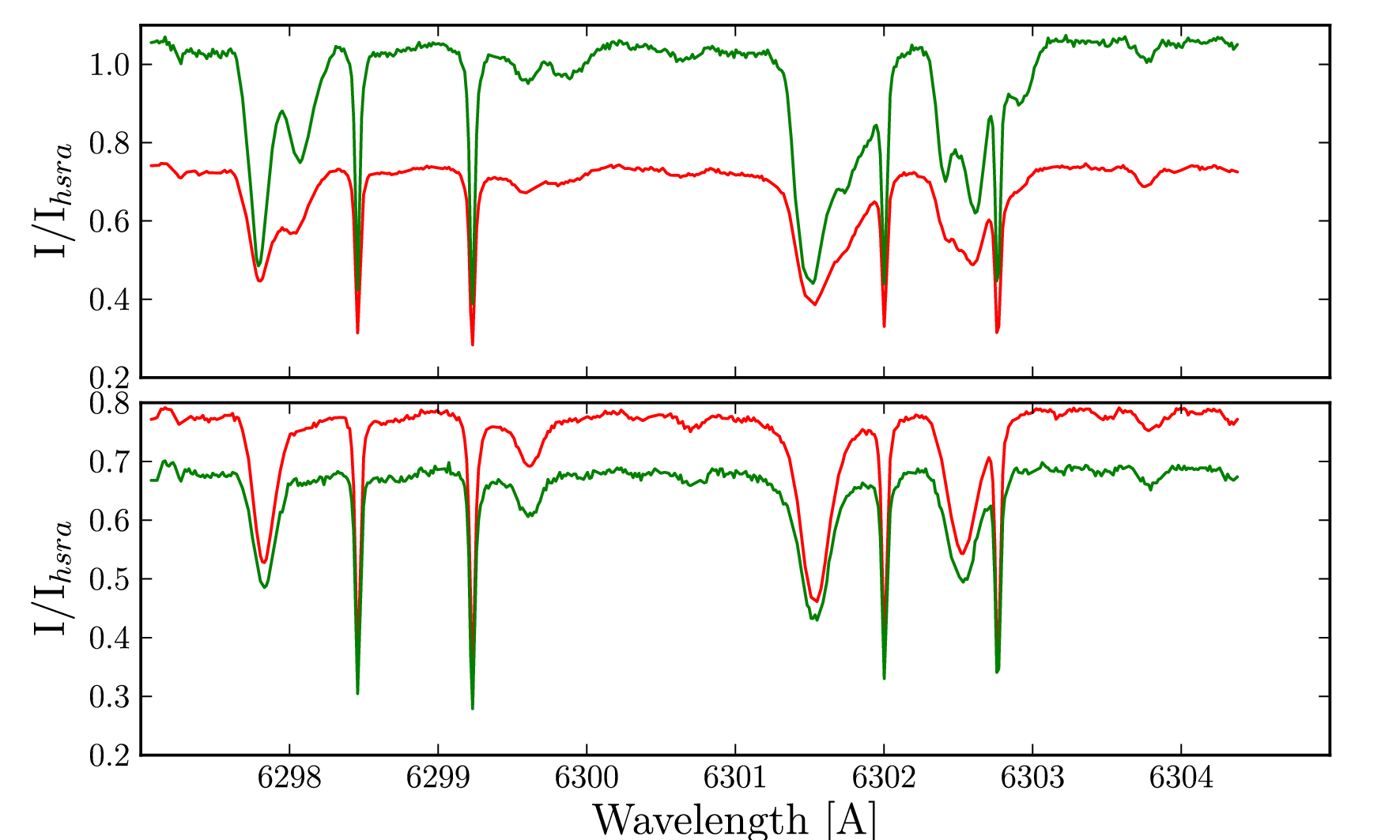}}
  \end{picture}
  \caption{Comparison of some line profiles from the unrestored (red) and restored (green) data. The top profile is from a penumbral filament protruding into the umbra of the spot in the top-left scan, the bottom profile is from a micropore from the bottom-left scan.}
  \label{fig:spectra}
\end{figure}

Since the image contrast is clearly significantly changed by the restoration, one would expect the spectra to show a similar behavior. In Fig.~\ref{fig:spectra}, we plotted the spectra for two locations in the data, one of a bright feature on the edge of the umbra and one of a dark feature in the midst of granulation.

Clearly, the change in the continuum level is significant, mostly due to the inherently large effect that the image degradation has in areas of enhanced contrast. In particular, the difference in the shape of the FeI line at 6302.5\AA\ in the top panel is noticable, and would lead to a significantly different result if it were to be inverted.

This effect has been explored in the spatially coupled inversion of Hinode data \citep{2012A+A...548A...5V} and work using this method, with the difference that here the deconvolution is employed independently and is not applied as part of an inversion. Future analysis of these data will show if and how accurately the spatially coupled inversion results can be reproduced using high-resolution ground based data.

\section{Conclusions}
We have demonstrated that using the latest-generation high-performance CMOS cameras, the restoration of image information from spectral data is possible, using information obtained from simultaneously acquired slit-jaw image data. The restored spectral data look very similar to image data restored from filtergraph data, except for the solar evolution inherently contained in them, and the high-resolution spectral information over a substantial wavelength range. 

The spatial resolution that is achieved is remarkably uniform even when the seeing conditions momentarily degrade. This is quite possibly due to the relatively long period of time that each point on the solar surface is sampled by the slit, so that the statistical fluctuations in the seeing are more likely to produce a high-quality realization than in the more usual higher cadence image data.

The method can be used on data that are actively scanned, but also on data that are passively scanned by the seeing, in which case it can be used to obtain high-cadence spectral data over a limited FOV.

\begin{acknowledgements}
The author gratefully acknowledges Hans-Peter D\"orr for his participation in the observing campaign at the SST, and the reduction of the spectral data that was used in this paper, and
Guus Sliepen for online technical assistance, without which the Adaptive Optics system would not have been operational at the time that the data were recorded.
This research has made use of NASA's Astrophysics Data System.
\end{acknowledgements}

\bibliography{ms}

\end{document}